\documentclass{WileyMSP-template}
\usepackage{physics}
\usepackage{graphicx}
\usepackage{amssymb} 
\newcommand{\figref}[1]{FIG.~\ref{#1}}

\begin{document}

\pagestyle{fancy}
\rhead{\includegraphics[width=2.5cm]{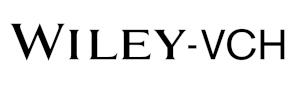}}

\title{
  Optimising motion-induced spin transfer
}

\maketitle

\author{Daigo Oue*}
\author{Mamoru Matsuo}

\begin{affiliations}
  Dr.~D.~Oue\\
  Instituto de Telecomunica\c{c}\~{o}es, Instituto Superior T\'{e}cnico, University of Lisbon, 1049-001 Lisbon, Portugal\\
  The Blackett Laboratory, Department of Physics, Imperial College London, Prince Consort Road, Kensington, London SW7 2AZ, United Kingdom\\
  Kavli Institute for Theoretical Sciences, University of Chinese Academy of Sciences, Beijing, 100190, China\\
  Email address: daigo.oue@gmail.com

  Prof.~M.~Matsuo\\
  Kavli Institute for Theoretical Sciences, University of Chinese Academy of Sciences, Beijing, 100190, China\\
  CAS Center for Excellence in Topological Quantum Computation, University of Chinese Academy of Sciences, Beijing 100190, China\\
  Advanced Science Research Center, Japan Atomic Energy Agency, Tokai, 319-1195, Japan\\
  RIKEN Center for Emergent Matter Science (CEMS), Wako, Saitama 351-0198, Japan\\
\end{affiliations}


\begin{abstract}
  In this paper, the spin transfer between two ferromagnetic insulators is studied.
  There is a narrow gap between the ferromagnetic insulators so that they are weakly interacting with each other.
  One of the ferromagnetic insulators is moving at a constant speed while the other is at rest; hence, the system is out of equilibrium.
  In the presence of the shearing motion, the interaction amplitude is periodically modulated at the Doppler frequency.
  A unitary transformation allows us to regard the periodic modulation of the interaction amplitude as an effective potential, which drives the spin transfer.
  The amount of the spin current is controlled by the spectral overlap and the carrier population difference between the two ferromagnetic media.
  If the spectra of the two ferromagnets are moderately broadened, the overlap in the spectral domain increases, enlarging the spin current.
  However, too much broadening spoils the spectral overlap and, hence, the spin current.
  This implies that there is an optimal condition for maximising the spin transfer.
\end{abstract}

\section{Introduction}
Tunnelling transport is a phenomenon that occurs between two closely situated media under the influence of an external bias. It is widely acknowledged as a pervasive nonequilibrium phenomenon that manifests across various fields of study.
For example, electron tunnelling transport between a metallic probe and a conducting sample realised electron tunnelling microscopy, an imaging method used to acquire ultra-high resolution at the atomic scale~\cite{binnig1983scanning}.
In addition to the conventional electron tunnelling, superconducting tunnelling junction has also been demonstrated~\cite{jaklevic1964quantum} and applied to high-sensitivity sensors for biological, chemical, and physical systems~\cite{fagaly2006superconducting,degen2017quantum,granata2016nano}. 
Engineering photon tunnelling transport between a sharply pointed tip and a sample enabled breaking diffraction limit and recording the subwavelength super-resolution optical image of the sample~\cite{pohl1984optical}.

\vspace{1em}
In a recently emerging field of spintronics, analogously to the preceding examples, spin tunnelling has been demonstrated.
This entails the transfer of spin between two closely situated media.
Various types of external control are adopted to induce imbalances between the two media.
Saitoh et al.~adopted microwave irradiation to drive the spin transfer between a ferromagnet film and a metallic medium~\cite{saitoh2006conversion}.
Uchida et al.~utilised the temperature difference between a ferromagnet and normal-metal wire(s) to trigger the spintronics analogue of the Seebeck effect (i.e.~current due to temperature gradient)~\cite{uchida2008observation}. 
Johnson and Silsbee induced chemical potential difference (spin accumulation) at the interface between a ferromagnetic and paramagnetic metal to generate spin current across the interface~\cite{johnson1985interfacial,johnson1988spin}.

\vspace{1em}
Here, we are going to theoretically investigate another possibility to drive spin transfer between two magnetic media.
Our proposal is inspired by a series of studies on noncontact friction, which is a momentum transfer between two closely positioned objects.
Typically, in the theoretical study of noncontact friction, the 'probe-sample' type setup is considered, where the sample is moving at a constant speed $v$, and the probe is placed near the sample.
As they are sufficiently close to each other, fundamental excitations in the probe and the sample can mutually interact, leading to linear momentum transfer between them, which is nothing but a frictional force without any physical contact.
There have been investigated various excitations mediating the momentum transfer: magnetostatic~\cite{saitoh2010gigantic,she2012noncontact,den2015spin}, electrostatic~\cite{stowe1999silicon,stipe2001noncontact,kuehn2006dielectric}, and radiative field~\cite{pendry1997shearing,zurita2004friction,volokitin2007near}.
In this work, we focus on spin angular momentum transfer rather than linear momentum transfer in such a 'probe-sample' setup.

\begin{figure}[tbp]
  \centering
  \includegraphics[width=.4\linewidth]{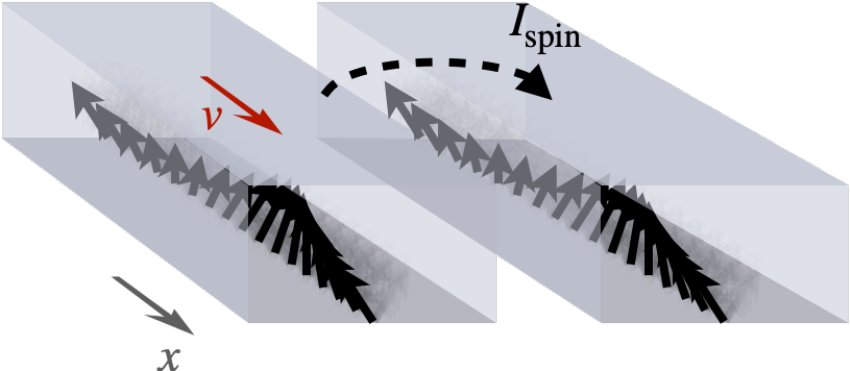}
  \caption{
    The schematic image of the setup that we will analyse in this work.
    We have two ferromagnetic insulators which are closely placed and, hence, interacting with each other.
    One is moving at a constant speed $v$ (red arrow).
    We shall focus on the fundamental excitations in the ferromagnetic insulators, magnons (indicated by solid black arrows), and analyse if the shearing motion drives the spin transfer (black dashed arrow).
  }
  \label{fig:setup}
\end{figure}
Our setup comprises two ferromagnetic insulators, which are magnetised in the $z$ direction, with a narrow vacuum gap in between (see \figref{fig:setup}).
One of the two ferromagnets is moving at a constant speed while the other is at rest.
We are going to investigate spin transfer from the moving to stationary magnets.
The amount of spin transfer can be evaluated by the time variation of total spins in the stationary medium,
\begin{align}
  I _ \mathrm{spin} = \int \pdv{\expval{S _ \mathrm{R} ^ {z}(x)}}{t} \dd{x} 
  = \frac{1}{i\hbar}\int \expval{\comm{S _ {\mathrm{R}} ^ {z}(x)}{H}} \dd{x},
  \label{eq:def:I _ spin}
\end{align}
where we introduced spin operators $S ^ {x,y,z}$ satisfying the angular momentum algebra (i.e.~$\comm{S ^ {j _ 1}}{S ^ {j _ 2}} = i \epsilon _ {j _ 1 j _ 2 j _ 3} S ^ {j _ 3}$),
and $H$ is the total Hamiltonian of the system given in the following.
Note that the average $\expval{\ldots}$ should be taken with respect to the total Hamiltonian.
Reminding that the vacuum gap separates two magnets composing our system and hence weakly interacting with each other, we can split the total Hamiltonian into three parts, $  H = H _ \mathrm{L} + H _ \mathrm{R} + H _ \mathrm{int}$, where we adopt the spin-exchange type tunnelling interaction between two magnets for the interaction part,
\begin{align}
  H _ \mathrm{int} = \int \hbar J _ \mathrm{int} S _ {\mathrm{L}} ^ + (x;v) S _ {\mathrm{R}} ^ - (x) \dd{x} + \mathrm{H.c.},
\end{align}
where we have written the coupling strength $J _ \mathrm{int}$ and defined spin flip operators, $S ^ + = S ^ x + iS ^ y$ and $S ^ - = \{S ^ +\} ^ \dagger$.
Note that we have explicitly written the sliding velocity in the argument of the spin operator for the left medium.
Note also that inclining the magnets may introduce a correction to the interaction strength $J _ \mathrm{int}$.
At the moment, we shall adopt the interaction picture and can give the bare Hamiltonian for each magnet later,
\begin{align}
  I _ \mathrm{spin} = \frac{1}{i\hbar}\int \expval{\comm{\widetilde{S} _ {\mathrm{R}} ^ {z}(x)}{H _ \mathrm{int}}} \dd{x} 
  = 2 J _ \mathrm{int} \Im \int \expval{\widetilde{S} _ {\mathrm{L}} ^ -(x;v) \widetilde{S} _ {\mathrm{R}} ^ +(x)} \dd{x},
  \label{eq:I _ spin (def)}
\end{align}
where operators with tilde symbols are evaluated in the interaction picture.
Here, the expectation value on the right-hand side $\expval{\widetilde{S} _ {\mathrm{L}} ^ - \widetilde{S} _ {\mathrm{R}} ^ +}$ should be assessed with special consideration on the fact that our system is pushed out of equilibrium by the persistent, constant motion imposed on the left medium.
Note that we shall omit the tilde symbols for brevity in the following although we keep working in the interaction picture.

\vspace{1em}
In our previous study~\cite{oue2022motion}, we phenomenologically employed an effective Hamiltonian capturing the nonequilibrium nature and showed how to evaluate the amount of the spin transfer~\eqref{eq:I _ spin (def)}; however, we did not delve into the mechanisms and reasons behind the system being driven out of equilibrium or derived the effective Hamiltonian.
In this article, we elucidate how and why the system is propelled out of equilibrium due to work done by the sliding motion (Sec.~2 backed by Sec.~6) and derive the effective Hamiltonian (Sec.~3).
Furthermore, after outlining how to quantify the amount of spin transfer following the Schwinger-Keldysh formalism (Sec.~4), we explore, within the lowest-order spin-wave approximation, the impact of spectral overlap between the two magnets on the spin transfer effect (Sec.~5).

\section{Work done by the sliding motion}
In this section, we describe how the forced sliding motion drives the system out of equilibrium.
According to the previous studies on noncontact friction, we can expect that there is finite linear momentum transfer between the two magnets, resulting in a force.
Working in the slow-velocity regime, we can assume that the force scales linearly with the velocity,
\begin{align}
  F _ \mathrm{f} = - \gamma v,
\end{align}
where $\gamma > 0$ is the friction coefficient, which can be microscopically determined, as we will show later.
As the external force $F _ \mathrm{ex}$ should be balanced with the frictional force to maintain the constant motion ($F _ \mathrm{ex} = -F _ \mathrm{f}$),
the work done by the sliding motion per unit of time is
\begin{align}
  W = F _ \mathrm{ex} \cdot v = \gamma v ^ 2 > 0.
\end{align}
This implies that work is perpetually done by the constant motion, causing the system to continuously receive energy and deviate out of equilibrium.
In other words, we continuously pump energy into the system through the noncontact friction.

\section{Temporal modulation of the tunnelling coupling by the constant motion}
\begin{figure}[tbp]
  \centering
  \includegraphics[width=.4\linewidth]{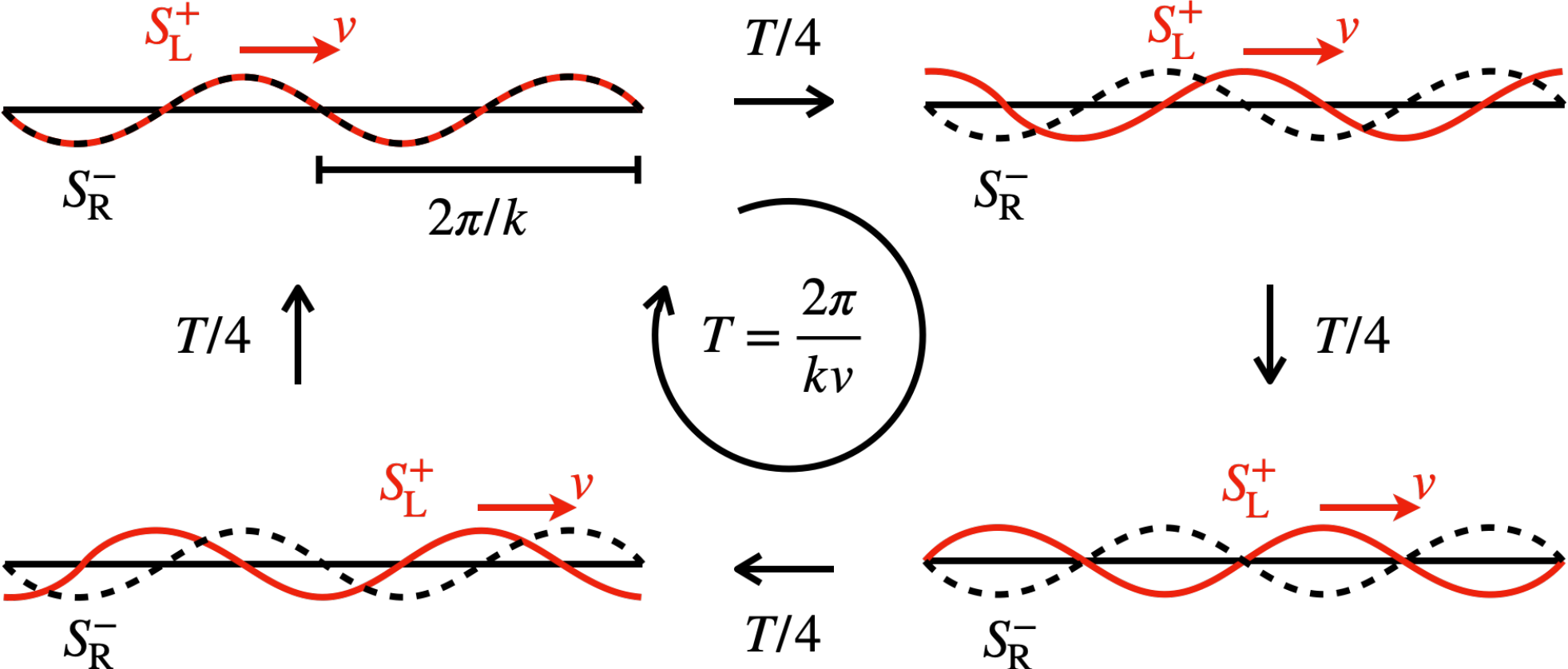}
  \caption{
    Cyclic modulation of the overlapping between left and right excitations.
    For the excitations with a wavelength $2\pi/k$, the overlap between the excitation is cyclically modulated with a period $T = 2\pi/(kv)$ due to the sliding motion.
  }
  \label{fig:cyclic}
\end{figure}
In this section, we shall study how the sliding motion affects the microscopic theory to confirm the system is indeed driven out of equilibrium.
As our left medium is moving at a constant velocity $v$, the spin operator in the laboratory frame is associated with the one in a reference frame where the left medium is at rest by applying the boost transformation.
Since we work in the slow-velocity regime, we can safely adopt the Galilean boost instead of the Lorentz one (we can deal with the relativistic correction just by replacing the Galilean transformation with the Lorentz one in the following) to write the spin operator for the moving medium,
\begin{align}
  S _ \mathrm{L}(x;v) = S _ \mathrm{L}(x-vt),
\end{align}
where the spin operator without the velocity in the argument, $S _ \mathrm{L}(x)$, is the one for the left medium at rest.

\vspace{1em}
Respecting the translation symmetry, we employ the Fourier representation,
\begin{align}
  S _ \mathrm{L} ^ + (x;v) &= \int S _ {\mathrm{L}k} ^ + e ^ {ik(x-vt)} \dd{k},
  \\
  S _ \mathrm{R} ^ + (x) &= \int S _ {\mathrm{R}k} ^ + e ^ {ikx} \dd{k}.
\end{align}
As a result, we can write the effective interaction Hamiltonian in the reciprocal space as
\begin{align}
  H _ \mathrm{int} = \int \hbar J _ {k}(t)  S _ {\mathrm{L}k} ^ + S _ {\mathrm{R}k} ^ - \dd{k} + \mathrm{H.c.},
  \quad
  J _ k (t) = J _ \mathrm{int} e ^ {-i \Delta \omega _ k t},
\end{align}
where the coupling constant is periodically modulated in time with the Doppler frequency $\Delta \omega _ k = kv$.
Note that this recovers the conventional tunnelling coupling Hamiltonian if we set $v=0$.
We can view the phase factor $e ^ {-i \Delta \omega _ k t}$ as a consequence of the cyclic modulation of overlapping between excitations in the left and right media (see \figref{fig:cyclic}).
Let us focus on excitations in the left and right media at a given wavenumber $k$.
The right medium is at rest; hence, the waveform of the excitation on the right does not change in time.
On the other hand, the left medium is moving at a constant velocity so that the excitation waveform alters in time.
Consequently, the overlapping is cyclically modulated in time with a period of $T = 2\pi/(kv)$.
The time dependence of the interaction Hamiltonian serves as the microscopic rationale behind the nonequilibrium nature of our system.

\vspace{1em}
Applying a unitary transformation, 
\begin{align}
  H \mapsto U H U ^\dagger - i\hbar U\pdv{t}U ^ \dagger,
\end{align}
the periodic modulation of the coupling strength can be reconsidered as an effective potential as follows.
We can choose
\begin{align}
  U = \exp \qty(+i \qty[\int \Delta \omega _ k S _ {\mathrm{L}k} ^ z \dd{k}] t)
\end{align}
to remove the time dependence of the coupling strength and get the conventional tunnelling coupling,
\begin{align}
  H _ \mathrm{int} \mapsto \int \hbar J _ \mathrm{int} S _ {\mathrm{L}k} ^ + S _ {\mathrm{R}k} ^ - \dd{k} + \mathrm{H.c.}.
\end{align}
In exchange for that, we have an effective potential on the left medium,
\begin{align}
  V := -i\hbar U\pdv{t}U ^ \dagger = 
  \int \Delta \omega _ k S _ {\mathrm{L}k} ^ z \dd{k}.
\end{align}
Therefore, the effective total Hamiltonian can be written as $H = H _ 0 + H _ \mathrm{int}$,
\begin{align}
  H _ 0 &= H _ \mathrm{R} + H _ \mathrm{L}',
  \label{eq:H _ 0}
  \\
  H _ \mathrm{L}' &= H _ \mathrm{L} + V,
  \\
  H _ \mathrm{int} &= \int \hbar J _ \mathrm{int} S _ {\mathrm{L}k} ^ + S _ {\mathrm{R}k} ^ - \dd{k} + \mathrm{H.c.},
  \label{eq:H _ int}
\end{align}
where the effects of sliding motion $V$ are included in the unperturbed part.
In our previous work~\cite{oue2022motion}, this Hamiltonian was phenomenologically introduced with the spin-wave approximation, and the motion-induced spin transfer was perturbatively evaluated.

\vspace{1em}
Here, we make a comment on the Galilean invariance.
No global coordinate translation can remove the system's motion, and we call for local translation to go over a reference frame where both magnets are at rest.
That is why we expect that the system no longer possesses the global Galilean invariance while it is still locally Galilean invariant.
Indeed, the Galilean invariance is broken due to the interaction between the two magnets $H _ \mathrm{int} \propto \int S _ \mathrm{L} ^ +(x-vt) S _ \mathrm{R} ^ -(x) \dd{x}$ as in the case of noncontact friction~\cite{reiche2022wading}
If the interaction is turned off, each system does not have any way to detect the motion, and the Galilean invariance is preserved.

\section{Motion-induced spin transfer}
In this section, we outline how to evaluate the motion-induced spin transfer with the perturbation theory with nonequilibrium Green's functions~\cite{oue2022motion}.
Working in the reciprocal domain, the quantity in question can be written as
\begin{align}
  I _ \mathrm{spin} 
  = 2 J _ \mathrm{int} \Im \int \expval{S _ {\mathrm{L}k} ^ - S _ {\mathrm{R}k} ^ +} \dd{k},
  \label{eq:I _ spin (reciprocal)}
\end{align}
where the expectation value should be calculated with the full Hamiltonian given in Equations~\eqref{eq:H _ 0} and \eqref{eq:H _ int}
Adopting the Schwinger-Keldysh formalism~\cite{schwinger1961brownian,keldysh1965diagram}, we can write the integrand in Equation~\eqref{eq:I _ spin (reciprocal)} as
\begin{align}
  \expval{\mathcal{T} _ \mathrm{C} S _ {\mathrm{R}k} ^ +(t _ 1 ^ +) S _ {\mathrm{L}k} ^ -(t _ 1 ^ -)}
  &= \expval{\mathcal{T} _ \mathrm{C} S _ {\mathrm{R}k} ^ +(t _ 1 ^ +) S _ {\mathrm{L}k} ^ -(t _ 1 ^ -) U _ \mathrm{C}} _ 0,
  \label{eq:integrand}
  \\
  U _ \mathrm{C} &= \mathcal{T} _ \mathrm{C}\exp (\frac{1}{i\hbar}\int _ \mathrm{C} H _ \mathrm{int}(t _ 2)\dd{t _ 2}),
\end{align}
where $\mathcal{T} _ \mathrm{C}$ is the time-ordering operator on the Schwinger-Keldysh contour, the superscript $+$ ($-$) at the argument stands for forward (backwards) branch, $\expval{\ldots} _ 0$ is the average with the unperturbed Hamiltonian $H _ 0$, and $\int _ \mathrm{C} \ldots \dd{t _ 2}$ is the integration over the Schwinger-Keldysh contour.
We expand $U _ \mathrm{C}$ in Equation~\eqref{eq:integrand} in powers of the tunnelling coupling $J _ \mathrm{int}$, retain up to the first order, and apply the Bloch-de Dominicis theorem and the Langreth theorem~\cite{haug2008quantum} to get
\begin{align}
  i\hbar ^ 2 J _ \mathrm{int}\int _ {-\infty} ^ {+\infty} \qty(
  \chi _ {\mathrm{R}k;12} ^ \mathfrak{R} \chi _ {\mathrm{L}k;21} ^ <
  + \chi _ {\mathrm{R}k;12} ^ < \chi _ {\mathrm{L}k;21} ^ \mathfrak{R}
  ) \dd{t _ 2},
  \label{eq:integrand-Langreth}
\end{align}
where $\chi _ {\mathrm{L(R)}k} ^ {\mathfrak{R},<,\mathfrak{A}}$ is the retarded, lesser, and advanced components of the nonequilibrium Green's functions for left (right) medium,
\begin{align}
  \chi _ {\mathrm{L(R)}k;12} = \frac{1}{i\hbar}\expval{\mathcal{T} _ \mathrm{C}S _ {\mathrm{L(R)}k} ^ + (t _ 1) S _ {\mathrm{L(R)}k} ^ - (t _ 2)} _ 0.
\end{align}
Note that the detailed analysis of \textit{charge} transfer phenomena between two adjacent media using non-equilibrium Green's functions has been carried out in detail by Meir and Wingreen~\cite{meir1992landauer}, where they summed up perturbations of all orders to derive a formula applicable even if the two media strongly interact with each other.
On the other hand, for \textit{spin} transfer phenomena between two adjacent media, it is sufficient to consider only the lowest order of perturbation, since it is typical for the two adjacent media to be weakly coupled~\cite{ohnuma2014enhanced,ohnuma2017theory,matsuo2018spin,kato2019microscopic,ominato2020valley}.
Applying the Fourier transformation to Equation~\eqref{eq:integrand-Langreth}, we can write
\begin{align}
  2i\hbar ^ 2 J _ \mathrm{int}\int \Im \chi _ {\mathrm{R}k\omega} ^ \mathfrak{R} \Im \chi _ {\mathrm{L}k\omega} ^ \mathfrak{R} \delta f _ {k\omega} \dd{\omega},
\end{align}
where we defined the distribution difference function with the retarded and lesser components of the nonequilibrium Green's functions,
\begin{align}
  \delta f _ {k\omega} = \frac{\chi _ {\mathrm{L}k\omega} ^ <}{2i\Im \chi _ {\mathrm{L}k\omega} ^ \mathfrak{R}}
  - \frac{\chi _ {\mathrm{R}k\omega} ^ <}{2i\Im \chi _ {\mathrm{R}k\omega} ^ \mathfrak{R}}.
  \label{eq:delta f}
\end{align}
From these considerations, we can write the total amount of the spin transfer as
\begin{align}
  I _ \mathrm{spin} 
  = 4 (\hbar J _ \mathrm{int}) ^ 2 \int \Im \chi _ {\mathrm{R}k\omega} ^ \mathfrak{R} \Im \chi _ {\mathrm{L}k\omega} ^ \mathfrak{R} \delta f _ {k\omega} \dd{k}\dd{\omega}.
  \label{eq:I _ spin - formula}
\end{align}
It is clear from this expression that the spin transfer process is the second order in the tunnelling coupling and proportional to the spectral overlap $\Im \chi _ {\mathrm{R}k\omega} ^ \mathfrak{R} \Im \chi _ {\mathrm{L}k\omega} ^ \mathfrak{R}$ and the distribution difference $\delta f _ {k\omega}$ between the left and right magnets.
In our previous study~\cite{oue2022motion}, we focused on the distribution difference~\eqref{eq:delta f} and studied how it plays a role in the motion-induced spin transfer within the spin-wave approximation.
In the following, we shall discuss the impact of the spectral overlap, a facet that was overlooked in our prior work, in the motion-induced spin transfer.
Note that we shall still keep the lowest-order spin-wave approximation to highlight the role of the spectral overlap.

\section{Role of Spectral Overlap}
To discuss the spectral overlap function, we apply the spin-wave approximation that captures the fundamental excitations in the ferromagnetic insulators.
The unperturbed Hamiltonian, which is responsible for the ferromagnets, is written in terms of bosonic operators according to the Holstein-Primakoff theory within the spin-wave approximation (e.g.~$S _ {\mathrm{L}k} ^ {+(-)} \approx \sqrt{2S _ 0} b _ {\mathrm{L}k} ^ {(\dagger)}$ and $S _ {\mathrm{L}k} ^ z \approx S _ 0 - b _ {\mathrm{L}k} ^ \dagger b _ {\mathrm{L}k}$ where $S _ 0$ is the magnitude of the localised spin, and $b _ {\mathrm{L}k}$ is a bosonic annihilation operator),
\begin{align}
  H _ \mathrm{R} &= \int \hbar \omega _ {k} b _ {\mathrm{R}k} ^ \dagger  b _ {\mathrm{R}k} \dd{k},
  \\
  H _ \mathrm{L}' &= \int \hbar (\omega _ {k} - \Delta \omega _ k) b _ {\mathrm{L}k} ^ \dagger  b _ {\mathrm{L}k} \dd{k},
\end{align}
where we substitute a conventional parabolic magnon dispersion $\omega _ k = D k ^ 2 + \omega _ 0$ with the Zeeman energy $\omega _ 0$.
Remind that $\Delta \omega _ k = kv$ is the Doppler-induced effective potential.
Note that we assumed the left and right media are made of the same material.

\vspace{1em}
As we have specified the unperturbed Hamiltonian, we can explicitly write the spectral function for each magnet.
Since we are working in the interaction picture, we can write
\begin{align}
  b _ {\mathrm{L(R)}k}(t) = e ^ {-i\frac{H _ \mathrm{L(R)}}{\hbar} t} b _ {\mathrm{L(R)}k} e ^ {+i\frac{H _ \mathrm{L(R)}}{\hbar} t} 
  = b _ {\mathrm{L(R)}k} e ^ {-i\omega _ {\mathrm{L(R)}k} t},
\end{align}
where we defined the magnon dispersion in the right magnet $\omega _ {\mathrm{R}k} = \omega _ k$ and the one in the left $\omega _ {\mathrm{L}k} = \omega _ k - \Delta \omega _ k$.
Therefore, the retarded component of the nonequilibrium Green's function can be explicitly written as
\begin{align}
  \chi _ {\mathrm{L(R)}k\omega} ^ \mathfrak{R} = \frac{2S _ 0/\hbar}{\omega - \omega _ {\mathrm{L(R)}k} + i\Gamma},
\end{align}
where we have phenomenologically introduced the spectral broadening factor $\Gamma$.

\begin{figure}[tbp]
  \centering
  \includegraphics[width=.5\linewidth]{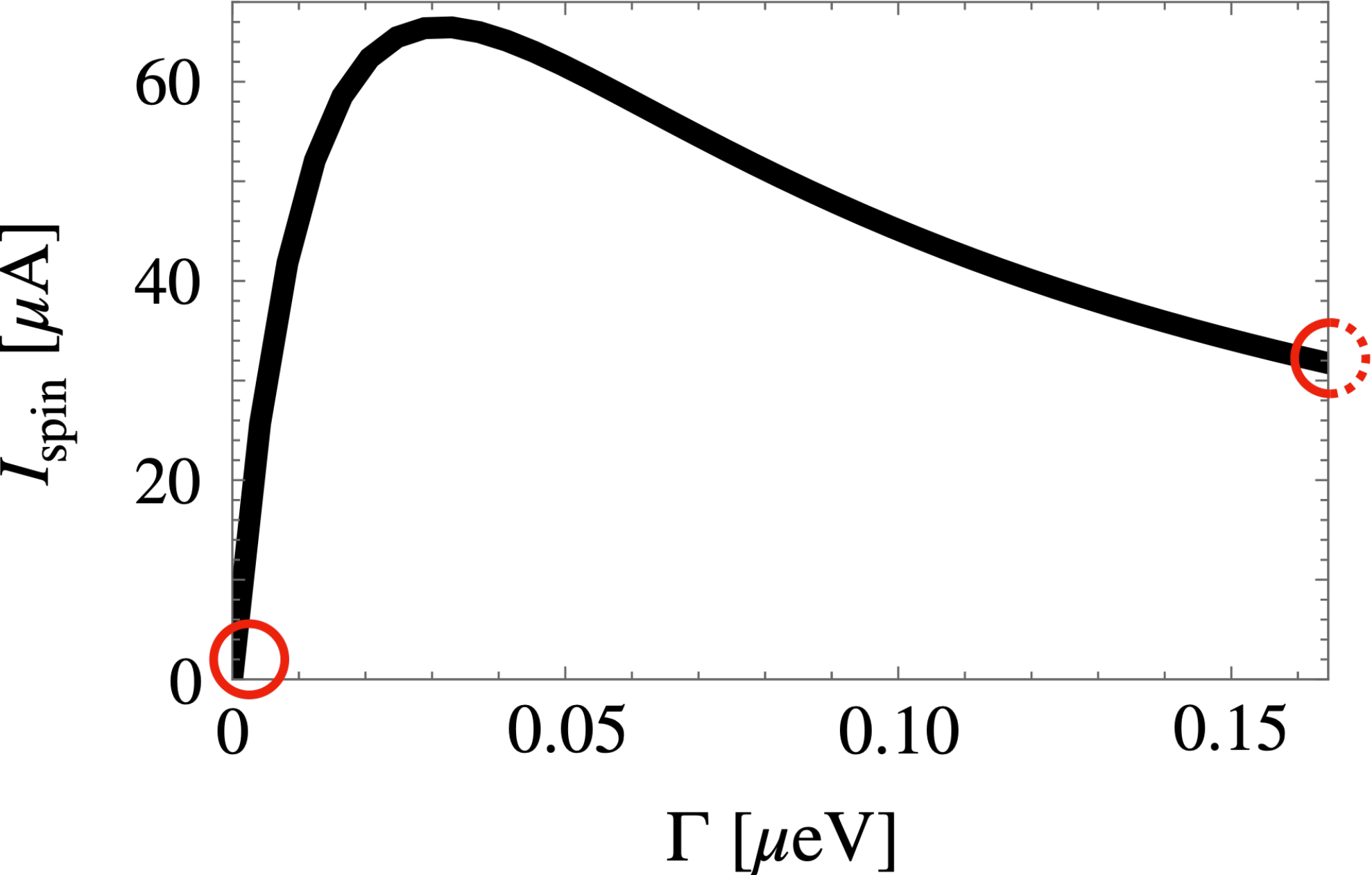}
  \caption{
    Effects of the spectral broadening on the spin transfer between the ferromagnetic insulators.
    In the 'pristine' limit ($\Gamma \rightarrow 0$), only $k=0$ channel is active, where the Doppler effect plays no role.
    In the 'dirty' limit ($\Gamma \rightarrow \infty$), the spectra are flattened, and the spectral overlap vanishes.
    The broadening enlarges the spectral overlap and the spin transfer; however, it spoils the transport if it is too much.
    We used the following parameters to draw this line plot:
    $D = 532\;\mathrm{meV\cdot\AA ^ 2};
    \omega _ 0 \approx 0.12\;\mathrm{meV} (B _ 0 = 1.0\;\mathrm{T});
    \hbar J _ \mathrm{int} = 50\;\mathrm{meV} \ll \hbar \omega _ k;
    v = 1.0\;\mathrm{m/s}.$
  }
  \label{fig:gamma}
\end{figure}

\vspace{1em}
Let us consider two extreme cases.
First, in the `pristine' limit ($\Gamma \rightarrow 0$), where the spectra are too sharp, they only overlap at $\omega _ {\mathrm{R}k} = \omega _ {\mathrm{L}k}$, where the spin carrier has no momentum $k = 0$, and the Doppler shift $\Delta \omega _ k = k\cdot v$ does not play any role; hence, there is no current induced by the sliding motion.
Second, in the `dirty' limit ($\Gamma \rightarrow \infty$) where the spectra are flattened, $\chi ^ \mathfrak{R} \rightarrow 0$, the spectral overlap will vanish, leading to the disappearance of the spin current (remind the spin current is proportional to the spectral overlap).
Thus, we can expect that there is some optimal condition where the induced spin current is maximised.

\vspace{1em}
In order to confirm the qualitative discussion, we numerically evaluated the spin transfer formula~\eqref{eq:I _ spin - formula}, sweeping the spectral broadening factor $\Gamma$.
In \figref{fig:gamma}, we show the amount of the motion-induced spin transfer $I _ \mathrm{spin}$ as a function of the spectral broadening.
We can verify that the spin current decreases as the broadening factor $\Gamma$ becomes too small and significantly large and that there is a peak in between.
In recent experiments~\cite{kosen2019microwave,pfirrmann2019magnons}, the linewidth of clean yttrium iron garnet, which is a ferromagnetic insulator considered in our numerical demonstration (\figref{fig:gamma}), reaches the order of $1\;\mathrm{MHz} \approx 4.1\;\mathrm{neV}$, which can be broadened with the introduction of impurities.

\section{Microscopic theory for the work done by the sliding motion}
In the preceding section, we employed the phenomenological description for the nonequilibrium nature of the system.
Here, we provide a microscopic theory for the work done by the sliding motion within the spin-wave approximation that we employed to evaluate the spin transfer in the previous section.
As the work done by the motion is provided by the momentum transfer between two magnets, we shall focus on the time variation of the momentum in the left magnet,
\begin{align}
  F _ \mathrm{f} = \int \hbar k \pdv{t} \expval{b _ {\mathrm{R}k} ^ \dagger b _ {\mathrm{R}k}} \dd{k} 
  = \int \hbar k \expval{b _ {\mathrm{L}k} ^ \dagger b _ {\mathrm{R}k}} \dd{k}.
\end{align}
Repeating the same procedure that we take to perturbatively evaluate the spin transfer, we can get
\begin{align}
  F _ \mathrm{f} = -4(\hbar J _ \mathrm{int}) ^ 2 \int \hbar k \Im \tilde \chi _ {\mathrm{L}k\omega} ^ \mathfrak{R} \Im \tilde\chi _ {\mathrm{R}k\omega} ^ \mathfrak{R} \delta f _ {k\omega} \dd{k}\dd{\omega}.
\end{align}
where we defined $\tilde\chi _ {\mathrm{L(R)}k\omega} ^ \mathfrak{R} = \chi _ {\mathrm{L(R)}k\omega} ^ \mathfrak{R}/(2S _ 0)$.
As the external force that keeps the left medium at a constant speed should be balanced, we can write
\begin{align}
  F _ \mathrm{ex} = -F _ \mathrm{f} 
  = 4(\hbar J _ \mathrm{int}) ^ 2 \int \hbar k \Im \tilde\chi _ {\mathrm{R}k\omega} ^ \mathfrak{R} \Im \tilde\chi _ {\mathrm{L}k\omega} ^ \mathfrak{R} \delta f _ {k\omega} \dd{k}\dd{\omega},
\end{align}
Assuming the sliding velocity $v$ is small, we expand the integrand and retain the first order in $v$ to obtain $F _ \mathrm{ex} = -F _ \mathrm{f} = \gamma v$ with a drag coefficient,
\begin{align}
  \gamma = 4(\hbar J _ \mathrm{int}) ^ 2 \int \hbar k ^ 2
  \Im \tilde\chi _ {k\omega} ^ \mathfrak{R} \Im \tilde\chi _ {k\omega} ^ \mathfrak{R} 
  n _ \mathrm{B}'(\omega _ k) 
  \dd{k}\dd{\omega},
\end{align}
where we defined $\tilde\chi _ {k\omega} ^ \mathfrak{R} = \eval{\tilde\chi _ {\mathrm{L}k\omega} ^ \mathfrak{R} }_ {v = 0} = \tilde\chi _ {\mathrm{R}k\omega} ^ \mathfrak{R}$, and $n _ \mathrm{B}'$ is the derivative of the Bose distribution function.
Note that we already performed the bosonisation and thus explicitly calculated the distribution difference where we get the Bose distribution.

\section{Discussion}
As we evaluated the drag force in the previous section, we can expand the integrand of the spin current formula~\eqref{eq:I _ spin - formula},
\begin{align}
  I _ \mathrm{spin} = \int \qty(\alpha _ 1 kv + \alpha _ 2 k ^ 2 v ^ 2 + \alpha _ 3 k ^ 3 v ^ 3 + \cdots) \dd{k}\dd{\omega},
  \label{eq:I _ spin - expansion}
\end{align}
to confirm the motion-induced spin transfer exhibits the parabolic dependence on the sliding velocity, $I _ \mathrm{spin} \approx \sigma _ \mathrm{s} ^ {(2)} v ^ 2$, in the slow-velocity regime ($v \ll 1$).
Note that we defined $\sigma _ \mathrm{s} ^ {(2)} = \int \alpha _ 2 k ^ 2 \dd{k}\dd{\omega}$.
As can be seen in the expansion~\eqref{eq:I _ spin - expansion}, the odd terms in $v$ are also odd in $k$, and the odd terms in $k$ do not contribute after the integration over $k$.
That is why the first order in $v$ does not contribute to the spin transfer, and the second order is the dominant contribution in the slow-velocity regime.

\vspace{1em}
The parabolic dependence on the sliding velocity implies we can obtain a second harmonic generation of spin current in our setup, which is forbidden in bulk due to symmetry.
Let us consider oscillating motion $v = v _ 0 \cos \Omega t$.
Even though the velocity is no longer constant, the present theory can still be applied adiabatically when the acceleration is sufficiently small $\dot{v} \ll 1$.
As our spin current is proportional to $v ^ 2$, we will generate the rectified and second-harmonic components of the spin current,
\begin{align}
  I _ \mathrm{spin} (t) \approx \sigma _ \mathrm{s} ^ {(2)} v _ 0 ^ 2 + \sigma _ \mathrm{s} ^ {(2)} v _ 0 ^ 2 \cos (2\Omega t),
\end{align}

\vspace{1em}
As for the experimental verification of our proposal, we can raise two potentially suitable setups. 
One possibility is to utilise a shear force microscope (SFM), which has been employed to measure noncontact friction experimentally~\cite{kuehn2006dielectric,saitoh2010gigantic,lee2015noncontact,wang2022non}.
In a SFM setup, we have a probe moving along a sample surface.
To verify our proposal with this setup, one may prepare a ferromagnetic tip, which hosts magnonic excitations, and move it along the surface of a ferromagnetic sample.
Another possibility may be provided by an analogous experiment where one makes use of spin-polarised electron beams~\cite{pierce1980gaas,guo1990further,batelaan1999optically,ahrendsen2023improved}.
When spin-polarised electrons pass nearby a substrate, they may interplay with the substrate with the spin-exchange type interaction.
Once the spin transfer is triggered in these setups or any other configurations, it can be detected by the inverse spin Hall measurement~\cite{ando2011inverse}.

\section{Conclusions}
In conclusion, we found that shearing motion drives spin transfer between two ferromagnetic insulators.
We clarified that the shearing motion in the presence of interaction between the two magnets breaks the global Galilean invariance and pumps energy into the system, driving it out of equilibrium and triggering transport phenomena.
Taking the nonequilibrium nature of the system, the amount of the spin transfer has been perturbatively assessed with nonequilibrium Green's functions.
The resulting spin current is controlled by the spectral overlap and the population difference between the two magnets.
We revealed the relevant (significant) spectral broadening enlarges (spoils) the spin current because it expands the spectrum overlap (flattens the spectra).

\medskip
\textbf{Acknowledgements}
D.O.~is supported by the JSPS Overseas Research Fellowship, by the Institution of Engineering and Technology (IET), and by Funda\c{c}\~ao para a Ci\^encia e a Tecnologia and Instituto de Telecomunica\c{c}\~oes under project UIDB/50008/2020.
This work was supported by the Priority Program of Chinese Academy of Sciences (Grant No.~XDB28000000) and JSPS KAKENHI (Grant No.~JP20H01863, No.~JP21H04565, and No.~JP21H01800).

\bibliographystyle{MSP}
\bibliography{%
  bib/all,%
  textbooks/textbook_all%
}
\end{document}